\documentstyle[12pt,openbib]{article}

\title{\bf Positronium atom scattering by H$_2$ in a coupled-channel
framework}

\author{P K Biswas$^{\$}$ and Sadhan K Adhikari$^{+}$ \\
$^{\$}$Departamento de F\'isica-IGCE, Universidade Estadual
Paulista\\  13500-970
Rio Claro, S\~ao Paulo, Brasil\\
$^{+}$Instituto de F\'isica Te\'orica, Universidade Estadual Paulista,\\
01405-900 S\~ao Paulo, S\~ao Paulo, Brasil}

\date{\today}
\begin{document}

\maketitle
\begin{abstract}

The scattering of ortho positronium (Ps) by H$_2$ has been investigated
using a three-Ps-state [Ps(1s,2s,2p)H$_2($X $^1\Sigma_g^+$)]
coupled-channel model and using Born approximation for higher excitations
and ionization of Ps and B $^1\Sigma_u^+$ and b $^3\Sigma_u^+$ excitations
of H$_2$. We employ a recently proposed time-reversal-symmetric nonlocal
electron-exchange model potential. We present a calculational scheme for
solving the body-frame fixed-nuclei coupled-channel scattering equations
for Ps-H$_2$, which simplifies the numerical solution technique
considerably.  Ps ionization is found to have the leading contribution to
target-elastic and all target-inelastic processes. The total cross
sections at low and medium energies are in good agreement with experiment.

\end{abstract}

\newpage

\section{Introduction}

With the availability of improved monoenergetic ortho positronium (Ps)
beam,
low-energy collision of exotic ortho Ps  atom with neutral
atoms and molecules is of interest in both physics and chemistry due
to its vast applicational potential \cite{1}. Recently, measurements of
total ortho-Ps scattering cross section from various atomic and molecular
targets (He, Ne, Ar, H$_2$, N$_2$, C$_4$H$_{10}$, and C$_5$H$_{12}$) have
been carried out at low and medium to high energies
 \cite{2,3,5}. Experimental investigations are also in progress for the
measurement of pick-off annihilation of ortho Ps from closed shell atoms
\cite{5x}. Because of the composite nature and
high
polarizability of Ps, a reliable theoretical description of Ps scattering
is far more complicated than electron  scattering \cite{12},
as
excitations and ionization of Ps are expected to play a
significant role for each target states. 
Besides if the target is vulnerable to excitations, prediction of total
cross section becomes extremely difficult as the number of scattering
channels grow as $N^2$, where $N$ excited states of both Ps and target are
considered.  
 
Due to internal charge and mass symmetry, Ps atom yields zero elastic and
even-parity transition potentials in the direct channel 
and Ps scattering
 is dominated mainly by exchange correlation
\cite{6,7,8}.
This
eventually complicates the convergence of a conventional coupled-channel
description \cite{6}. For Ps scattering with molecular target we have
additional
complication due to the presence of the nuclear degrees of freedom. The
presence of three charge centers complicates the solution scheme of the
coupled-channel equations. The degree of complication can be realized from
the mathematical analysis of ref. \cite{14} for Ps formation in
positron-hydrogen-molecule scattering. Because of mathematical
complication, this approach has not been further pursued in numerical
analysis and there exists no successful calculation of Ps formation in
positron-hydrogen-molecule scattering. We recall that the mathematical
complication of a dynamical coupled-channel study of Ps formation in
positron-hydrogen-molecule
scattering is similar to that of Ps-H$_2$ scattering.

Here we undertake a theoretical study of Ps scattering by H$_2$.  We
employ a hybrid approach   of treating the molecular
orientational
dependence as parameters in the coupled equations and then perform the
partial-wave expansion \cite{lane}.  The resulting one-dimensional
equations are
solved at the equilibrium nuclear separation by the matrix-inversion
technique and the partial-wave cross sections are numerically averaged
over molecular orientations.  The present approach makes the Ps-H$_2$
scattering problem easily tractable.

Recently, we suggested a regularized electron-exchange model potential
\cite{12,7}
and
demonstrated its effectiveness in exchange-dominated Ps scattering by
performing quantum coupled-channel calculations 
using the ab initio framework of close-coupling method for both simple and
complex targets
\cite{6,10,11,13,13x}. In our initial
calculations we used a nonsymmetric model-exchange potential for Ps
scattering by H \cite{10}, He \cite{7} and H$_2$ \cite{12,11} and obtained
reasonably good agreement with experiment on He and H$_2$.  In a
subsequent application of Ps scattering by H, it was found that a
time-reversal symmetric form of the exchange potential leads to far
superior result than the nonsymmetric form both in qualitative and
quantitative agreement with accurate variational calculations on
H \cite{6}. The symmetric potential also  led to very good results
for low-energy cross sections 
for  Ps
scattering by He, Ne, and Ar \cite{13a} in excellent agreement with
experiment \cite{5}.  

In view of the above  we reinvestigate
the problem of Ps scattering by H$_2$  using the symmetric
exchange potential employing the three-Ps-state [Ps(1s,2s,2p)]
coupled-channel 
model for elastic and
Ps(2s,2p) excitations.
We solve the
coupled-channel equations by the above  scheme  and  report
partial cross sections for Ps(1s,2s,2p) excitations.
We also calculate  cross sections for Ps  excitations to  $6\ge n\ge 3$
states  and Ps ionization using  the first Born approximation.  A
target-elastic total cross section is calculated by adding the above
partial cross sections. We  use  the present symmetric exchange potential 
to calculate the target-inelastic Born cross sections for excitation to  
B $^1\Sigma_u^+$
and b $^3\Sigma_u^+$ states  of H$_2$ 
 by Ps impact.   We present  a total Ps-H$_2$ cross section
by adding the above target-elastic and target-inelastic results.

In section 2 we present the theoretical formulation in the body-frame 
fixed-nuclei approximation. In section 3 we present the
numerical results. Finally, in section 4 we present a summary of our
findings.

\section{Theoretical Formulation}

The total wave function ${ \Psi}$
of the Ps-H$_2$ system is expanded in terms of
the Ps and H$_2$ quantum states as 
\begin{eqnarray} \label{zz}
{ \Psi}({\bf
r_0, r_1,r_2,x;R
})&=& {\cal A}
\sum_{a,b}\biggr[ F_{ab}({\bf s_0})\varphi_a ({\bf t_0
})\psi_b({\bf r_1,r_2;R}) \biggr]
\end{eqnarray}
where ${\bf s_0= (x+r_0)}/2$, ${\bf t_0= (x-r_0)}$, 
$\bf x$ ($\bf r_0$) is the coordinate of the positron (electron)
of the Ps atom, 
$\bf r_1$ and $\bf r_2$ are the
coordinates of the
electrons of H$_2$, $2{\bf R}$ is the internuclear separation of H$_2$,
$\varphi_a$  the wave function of 
Ps, $\psi_b$  the wave function of H$_2$,    
 with $b$ denoting the electronic configuration  of H$_2$ and $a$
denoting the quantum state of  Ps. 
Here ${\cal A}$ denotes antisymmetrization
with respect to Ps- and target-electron coordinates and 
 $F_{ab}$ is the
continuum orbital of Ps with respect to the target.  
The spin of the positron is
conserved in this process and the exchange profile of the Ps-target system
is analogous to the corresponding electron-target system \cite{13}.

Primary complication arises in the coupled-channel study
in retaining both the
summations over Ps and target states in the coupling scheme. As we shall
see
later that the contribution of the individual target-inelastic  channels 
to cross section 
is one order   small compared to that of target-elastic channels. So we
exclude
the
summation over target states from the coupling scheme and treat cross
section for target excitations 
separately using Born approximation.  
In this  work we are mostly
interested to see whether the  model can accommodate  the measured
total
cross
sections of Ps-H$_2$ scattering. So we retain the Ps(1s,2s,2p)
excitations in the
coupled-channel model and treat the remaining excitations by Born
approximation. 
The  first Born calculation with present regularized 
exchange leads to  results  reasonably close to coupled-channel
calculation at medium energies. Hence it is expected that this
calculation should yield a fairly good picture of the total cross section
except near the Ps-excitation thresholds.

Projecting the
resultant Schr\"odinger equation 
on the final Ps and target   states and
averaging over   spin, the resulting
momentum-space Lippmann-Schwinger scattering equation in the body-frame 
representation 
for a particular
total electronic spin state $S$ can  be
written as 
\begin{eqnarray}\label{1} &{}&f^{S}_{a ' a}({\bf k',k;R})={\cal
B}^{S}_{a 'a}({\bf k ',k;R})  - \frac{1}{2\pi^2}\sum_{a
''} \int \makebox{d}{\bf k ''}\frac {{\cal B}_{a' a ''}^S({\bf
k',k'';R}) f_{a'' a}^S({\bf k'',k;R}) } {
k_{a ''}^2/4
-k''^2/4+\makebox{i}0} \end{eqnarray}
where $f_{a'a}^S$ are the scattering amplitudes, and ${\cal B}^S _{a
'a}$  the corresponding Born potentials.
$k_{a ''} 
$ is the on-shell
 momentum of Ps in the intermediate channel $a ''$.
We use
units $\hbar = m = 1$ where $m$ is the electronic mass. For Ps-H$_2$
target-elastic scattering  there is only one scattering equation (\ref{1})
corresponding to total electronic spin $S=1/2$.  
The input potential  of (\ref{1}) is
given by 
\begin{equation}\label{2}
 {\cal B}^{1/2}_{a'a}({\bf k',k;R}) = B^{D}_{a'
a}({\bf
k',k;R})- B^{E}_{a' a}({\bf k',k;R})
\end{equation}
where $B^D$ is the direct Born potential and $B^{E}$ is the 
 exchange potential.  As in the ground electronic state $X^1\Sigma_g^+$
of H$_2$, the total electronic spin is zero, there will contribution of
one target electron to $B^{E}_{a' a}({\bf k',k;R})$ \cite{15}.

For the electronic ground state X $^1\Sigma_g^+$ of H$_2$
we use the  wave function of the form 
  $\psi_b({\bf r_1,  r_2;  R})$ $ \equiv 1\sigma^{(b)}
_g(1)\-1\sigma^{(b)} _g(2)
= N^2U_0({\bf r_1;        
 R})U_0({\bf r_2;  R}),$ where $N=[2(1+{\cal T})]^{-1/2}$ with ${\cal T}
$  the  overlap integral \cite{16aa} and  
$U_0 ({\bf r;  R})= (\delta^
3/\pi)^{ 
1/2} [\exp(-\delta                                                            
|{\bf r - R}|)+\exp (-\delta|{\bf r+ R}|)]$ with $\delta =1.166$
 \cite{16aa,16}.  For Ps we use the exact wave functions, e.g., for 1s
state
$\varphi(r) = \exp(-\beta r)/\sqrt{8\pi }$ with $\beta = 0.5$.  For the
excited states of H$_2$ we take the configurations $
\psi_{b'}({\bf r_1,  r_2;  R}) \equiv [1\sigma^{(b')}_g(1)
1\sigma^{(b')}_u(2) \pm 1\sigma^{(b')}_g(2) 1\sigma^{(b')} _u(1)]$, where
+
($-$) corresponds to
the spin singlet (triplet) state. All wave functions of H$_2$ are taken
from ref. \cite{16}.

The direct Born potential for Ps transition from state $a$ to $a'$  and
H$_2$ transition from $b$ to $b'$ 
can be rewritten in the following  convenient
factorized 
form \cite{18}:
\begin{eqnarray}\label{5}
{\cal B}^{D}_{a'b' \leftarrow ab} ({\bf k_f}, {\bf k_i; R})
& = & 
\frac {4}{Q^2} \int d{\bf t}
\varphi_{a'}^* ({\bf t}) [e^{i{\bf Q .  t} /2}- e^{-i{\bf Q .  t} /2}]
\varphi_a({\bf t}) \nonumber\\
& \times & \int d{\bf r_1}  d {\bf r_2}
 \psi_{b'}^*({\bf r_1,   r_2;  R})\left[ 2 \cos
({\bf Q. R}) - \sum_{n=1}^2e^{i{\bf Q.  r_n}}\right]
\psi_b({\bf r_1,   r_2; R}). 
\end{eqnarray}

Next we describe the electron-exchange model potential for Ps-H$_2$
scattering. We develop the model exchange potential from the following
term \cite{12,11}:
\begin{eqnarray}
{\cal B}^{E}_{a'b'\leftarrow ab} ({\bf k}_{\bf f}, {\bf k}_{\bf i};{\bf
R}) & = & 
- \frac{1} {\pi} \int d{\bf x} d{\bf r}_0 d{\bf r_1} d{\bf r_2}
e^{-i {\bf k_f . ( x +  r_1)}/2}
\varphi_{a'}^* ({\bf x-  r_1}) \psi_{b'}^* ({\bf r_0,   r_2; 
R})
\frac{1}  {|{\bf r_0 -  r_1}|} \nonumber \\
&\times &\psi_{b}({\bf r_1,   r_2;  R}) 
\varphi_a ({\bf x-  r_0}) e^{i {\bf k_i . ( x +  r_0)}/2}.
\label{3}
\end{eqnarray}
After removing the nonorthogonality of the initial and final wave
functions of (\ref{3}) and  some straightforward simplification 
the model exchange potential for a general target-elastic transition 
becomes \cite{12,7}
\begin{eqnarray}
{\cal B}^{E}_{a'b\leftarrow ab} ({\bf k_f}, {\bf k_i};{\bf R}) &
= & 
 \frac{4(-1)^{l+l'}} {
(k_f^2+k_i^2)/8+(2\delta_{b(g)}^2+\beta_a^2+\beta_{a'}^2)/2}
 \int d{\bf t}
\varphi_{a'}^* ({\bf t})  e^ {i {\bf Q .  t}/2} \varphi_a ({\bf
t})\nonumber\\
&\times &
\int d{\bf r_2  }d{\bf r_0} \psi_{b}^*({\bf r_0,  r_2;  R})
e^{i {\bf Q .  r_0}} \psi_{b}({\bf r_0,  r_2;  R}) \\
&
= &
 \frac{4(-1)^{l+l'}} {
(k_f^2+k_i^2)/8+(2\delta_{b(g)}^2+\beta_a^2+\beta_{a'}^2)/2}
 \int d{\bf t}
\varphi_{a'}^* ({\bf t})  e^ {i {\bf Q .  t}/2} \varphi_a ({\bf t})
\nonumber\\
&\times &
\int d{\bf r_2  } 1\sigma_g^{(b)}({\bf r_2})1\sigma_g^{(b)}({\bf r_2})
\int d{\bf r_0}    1\sigma_g^{(b)}({\bf r_0}) e^ {i {\bf Q .  r_0}}
1\sigma_g^{(b)}({\bf r_0})
\label{4}
\end{eqnarray}
where $l$ and $l'$ are the angular momenta of the initial and final
Ps states and  ${\bf Q=  k_i -k_f}$. 
Here $\delta_{b(g)}$ and $\beta_a$ are
the parameters of the H$_2$ and Ps wave functions in the initial state. 
The parameter $\beta_{a'}^2$ corresponds
to the final-state binding energy of the Ps atom and is taken as zero
while considering exchange for the Ps ionization channel. For
target-inelastic processes, following the prescription outlined in refs. 
\cite{12,7}
the
exchange potential takes the form 
\begin{eqnarray}
{\cal B}^{E}_{a'b'\leftarrow ab} ({\bf k_f}, {\bf k_i};{\bf R}) 
&
= &
 \frac{4(-1)^{l+l'}} {
(k_f^2+k_i^2)/8+(\delta_{b(g)}^2+\delta_{b'(u)}^2+\beta_a^2+\beta_{a'}^2)/2}
 \int d{\bf t}
\varphi_{a'}^* ({\bf t})  e^ {i {\bf Q .  t}/2} \varphi_a ({\bf t})
\nonumber\\
&\times &
\int d{\bf r_2  } 1\sigma_g^{(b')}({\bf r_2})1\sigma_g^{(b)}({\bf r_2})
\int d{\bf r_0}    1\sigma_u^{(b')}({\bf r_0}) e^ {i {\bf Q .  r_0}}
1\sigma_g^{(b)}({\bf r_0}).
\label{4x}
\end{eqnarray}
In (\ref{4x}) the additional indices $b$ and $b'$ are introduced on the
molecular orbitals 
$1\sigma_g$ and $1\sigma_u$ to distinguish the initial and final states. 
The
model exchange 
potentials  (\ref{4}) and (\ref{4x}) may be considered as a generalization
of a similar 
potential suggested by  Rudge \cite{17} for electron-atom
scattering.
The nonlocal and time-reversal symmetric exchange
potential (\ref{4}) has a very convenient  form and is expressed as 
a product of form factors of Ps and H$_2$.
Both the direct and exchange amplitudes have been factored out in terms of
Ps and target ``form-factors"  leading to a substantial simplification of
the theoretical calculation. 
 In our previous study 
of Ps-H$_2$ scattering \cite{12,11}
the prefactor of the exchange potential was not time-reversal symmetric.
In Eq. (\ref{4})  we have restored time-reversal symmetry as in ref. 
\cite{7} which is found to provide significant improvement in the results. 
Although,  for 
electron-molecule scattering several  model potentials are found in
the literature
\cite{16a,20}, there is no other  convenient model exchange potential 
for Ps scattering.

\section{Numerical Procedure and Results}

In the body-frame  calculation the coupled-channel equations are
solved at the  equilibrium nuclear separation $2R_0=1.4a_0$.
The polar and azimuthal
angles
$\theta_R$
and $\phi_R$ of $\bf R$
are taken as  parameters in the coupled equations \cite{lane}.
This reduces (\ref{1}) to the following form 
\begin{eqnarray} &{}&f^{R_0,\theta_R,\phi_R}_{a ' a}({\bf k',k})={\cal
B}^{R_0,\theta_R,\phi_R}_{a 'a}({\bf k ',k})  - \frac{1}{2\pi^2}\sum_{a
''} \int \makebox{d}{\bf k ''}\frac {{\cal B}_{a' a
''}^{R_0,\theta_R,\phi_R}({\bf
k',k''}) f_{a'' a}^{R_0,\theta_R,\phi_R}({\bf k'',k}) } {
k_{a ''}^2/4
-k''^2/4+\makebox{i}0}. \label{7}\end{eqnarray}
After standard partial-wave projection the three-dimensional coupled
 scattering equations (\ref{7}) are first reduced to coupled
one-dimensional
integral equations in momentum space. The one-dimensional equations  
are then  discretized by Gauss-Legendre 
quadrature rule and solved by the matrix inversion technique. 
A maximum of forty points
are used in the discretization of the integrals. 
The discretized coupled-channel
equations are solved for eight to ten discrete values each of
polar and azimuthal angles  $\theta_R$
and $\phi_R$. This leads to a   convergence of cross sections  up to
three significant figures. For targets with more charge asymmetry and for
polar molecules we expect that more points will be required for
angular averaging.  
The present   averaging  amounts to solving
the coupled set of scattering equations sixty four
to hundred 
times. 
Although this procedure increases the computational (CPU)
time, mathematical complications of tedious (and untractable)
angular-momentum
analysis \cite{14} are thus replaced by a tractable and
convenient calculational scheme.
Finally, 
the partial-wave    cross sections are numerically averaged
over molecular orientation using Gauss-Legendre quadrature points for both
polar
and azimuthal angles  $\theta_R$ and $\phi_R$.
Maximum number of partial waves included in the calculation is 12.
Contribution of higher partial waves to cross section is included by
corresponding Born terms.  These  Born  cross sections  are also
numerically averaged over all molecular
orientations in a similar fashion.

In figure 1, we show the angle-integrated target-elastic cross sections
for elastic, Ps(2s+2p),  Ps($3\le n\le 6$) excitations and
ionization of  Ps.  
As expected and observed
in previous calculations \cite{6,7,10,11} with other targets, the
contribution of
the Ps ionization channel to the cross section plays a dominant role from
medium to high energies as can be seen in figure 1. 
The detailed 
angle-integrated partial
cross sections of the  Born and three-Ps-state calculation 
are tabulated in table 1. Near 10 eV, 20 eV and 30 eV  the total Born
cross sections for Ps(1s,2s,2p) 
are found to be nonconvergent by 28$\%$, 9$\%$ and 4$\%$, respectively. 
Near 10 eV, 20 eV and 30 eV  the total Born
cross sections for Ps($n=2$) excitations 
are found to be nonconvergent by 18$\%$, 6$\%$ and 3$\%$, respectively.
At 60 eV the Born and three-Ps-state results are essentially identical. 
The nonconvergence of Ps($n\ge 3$) excitations calculated using Born
approximation are expected to lie within the limit set by Ps($n=2$) cross 
sections above. 
It is expected from experience that the ionization cross section
calculated using Coulomb Born will be more converged than the Ps($n=2$)
Born cross sections.

\vskip .2cm {Table 1: Ps-H$_2$ partial cross sections in units of $\pi
a_0^2$ at different positronium energies using the Born approximation and 
  three-Ps-state calculation} \vskip .2cm 
\begin{centering}
\begin{tabular} {|c|c|c|c|c|c|c|c|c|} 
\hline E (eV)& Ps(1s)&Ps(1s) & Ps(2s)&Ps(2s)  &Ps(2p)&Ps(2p)  & Ps($n\ge
3$) & Ps-ion \\
 &  Born & 3-St &Born & 3-St &Born &3-St&Born &Born\\
  \hline
 0.068&23.72 &3.79  &   & &        &       & & \\
 0.612&17.56&3.16  &       & &     &       & & \\
 1.45&11.84&2.47 &            & & &       & & \\
 3&6.71&1.60  &            &      & &  & & \\
 4& 5.03&1.17  &            &       && & & \\
 5&3.94&0.81  &            &       & & & &\\
 6 & 3.18&1.01&0.098&0.23 &3.46&2.28  &0 & 0 \\
 7 & 2.63&1.05&0.101&0.18 &3.60&2.52 & 1.28 &0.11  \\
 8 & 2.22&1.03& 0.092&0.14 &3.36&2.51 & 1.43 & 1.15 \\
 10 & 1.65&0.94&0.072&0.090 &2.80&2.27  &1.29 & 3.03 \\
 12.5 &1.21& 0.81&0.052&0.068 & 2.24&1.93& 1.05 & 4.34 \\
 15 & 0.92&0.68&0.041&0.054 &1.85&1.65 & 0.87 & 4.94 \\
 20 & 0.59&0.49&0.026&0.035 &1.34&1.25  &0.63 & 5.18 \\
 25 & 0.40&0.36&0.017&0.023 &1.04&0.99  &0.49 &4.93  \\
 30 & 0.29&0.27&0.012&0.017 &0.84&0.81  &0.40 &4.55  \\
 40 & 0.17&0.16&0.007&0.009 &0.60&0.58  & 0.28& 3.80 \\
 60 & 0.07&0.07&0.003&0.004 &0.37&0.37  & 0.17& 2.72 \\
\hline
\end{tabular}

\end{centering}

\vskip 0.3cm

Now we concentrate on some target inelastic processes. Each target
inelastic transition is accompanied by elastic, excitation and ionization
of Ps and hence by an infinite number of possibilities. Here to account
for target inelastic processes we consider Ps($n=1 \to 6$) discrete
excitations and ionization of Ps using the first Born approximation.
Contribution of higher discrete excitations of Ps are expected to be
insignificant and are neglected in the calculation. We calculate
the cross sections for the inelastic transition Ps(1s) + H$_2$(X
${}^1\Sigma^+_g) \to $ Ps* + H$_2$ (B ${}^1\Sigma_u^+$) where Ps*
represents the ground and the excited states of the Ps atom.  These Born
cross sections are also calculated at a equilibrium internuclear
separation $2R_0=1.4a_0$ and finally averaged over angular orientations of
the target. In figure 2 we display the Born contribution of partial cross
sections for transition to different Ps states while the hydrogen molecule
is excited to the H$_2$ (B ${}^1\Sigma_u^+$) state. The total cross
section summing the different contributions is also shown for this
target-inelastic process. When compared with the corresponding total cross
section calculated with the nonsymmetric exchange potential \cite{12} (not
shown in figure 2)  we find marginal change at low energies (below 20 eV)
and basically no change at medium to high energies.  This is quite
expected as the effect of exchange dies out at higher energies and these
cross sections are controlled by the direct Born potentials.

Next we calculate the cross sections for the inelastic transition Ps(1s) +
H$_2$(X ${}^1\Sigma^+_g) \to $ Ps* + H$_2$(b ${}^3\Sigma_u^+$) using the
first Born approximation, where Ps* 
represents the ground and the abovementioned excited states of the Ps
atom. 
These cross sections are again calculated for equilibrium 
internuclear separation $2R_0=1.4a_0$ and  averaged over 
molecular orientations.
In figure 3 we
display the contribution of partial cross sections for transition to
different Ps states while the hydrogen molecule is excited to the H$_2$
(b ${}^3\Sigma_u^+$) state. The total cross section summing the different
contributions is also shown for this target-inelastic process. 
There is
significant change when we compare this total cross section calculated
with the symmetric exchange potential with our previous result \cite{12}
obtained with the nonsymmetric exchange potential. This is quite expected 
as this process is purely exchange dominated. The qualitative and
quantitative differences between the two total cross sections reveal 
the importance of using the symmetric exchange potential.

In figure 4, we exhibit the  total cross section obtained from
target-elastic processes (Ps excitation up to $n=6$ and Ps ionization)
plus target excitations to B ${}^1\Sigma_u^+$ and b $^3\Sigma_u^+$ states
\cite{12,11} and compare with the total cross section measurements of
ortho-Ps scattering from H$_2$ \cite{2,5}.  The agreement with the
experimental cross sections is highly encouraging. Here, we have
considered only two lowest target-inelastic processes and their combined
effect on the total cross section. This gives an indication that the
inclusion of remaining important target-inelastic cross sections might
give a better agreement with measurement.

However, the present theoretical peak in total cross section is shifted to
a lower energy compared to experiment.  This trend was also found in our
calculation of Ps-He scattering \cite{6}. In the energy range of 7 to 15
eV, the  total cross section overestimates the measured data. This
is due to the fact that Ps($n\ge 3$) excitations and ionization have been
treated in the first Born approximation framework. The neglect of other
target-inelastic channels and the first Born calculation for higher
excitations and, in particular, ionization of Ps are supposed to be
responsible for the shift in the theoretical peak.  A dynamical
calculation for higher Ps excitations and ionization should reduce the
theoretical total cross section in the intermediate energy region, while
the inclusion of further target-inelastic channels will increase the cross
section for medium to higher energies. These two effects are expected to
shift the peak in the total cross section to higher energies and lead to a
better agreement with experiment. 

A previous first Born description of
Ps-H$_2$ scattering \cite{19} with Born-Oppenheimer exchange \cite{22} 
led to unphysically large cross section  at low energies. 
The poor performance of that
 scheme in Ps scattering compared to electron 
scattering is due to the fact that in the absence of the  direct
potential (zero for elastic scattering), the 
Born-Oppenheimer exchange potential solely determines the
cross section. This clearly shows the very unrealistic nature of the
Born-Oppenheimer exchange potential at low energies.

In previous studies \cite{6,7,13} of Ps scattering we introduced a
parameter $C$ in the
exchange potential for obtaining a more quantitative fit with experiment
essentially by replacing the denominator term in (\ref{4})  by 
$(k_f^2+k_i^2)/8+C(\delta_a^2+\delta_{a'}^2+\beta_b^2+\beta_{b'}^2)/2.$
In the original form the constant $C = 1$ and we have used this value in
this study. However, it can be varied slightly from
unity to obtain a precise  fit of a low-energy scattering observable
(experimental or variational), as has been done in some applications of
model potentials \cite{20}. A variation of $C$ from unity leads to a
variation of the average values for square of momenta \cite{7,10}, which 
were taken as
the binding energy parameters ($\delta^2, \beta^2$ etc.)
in the
expression for the denominator in   (\ref{4}). This variation, in turn,
tunes the
strength of
the exchange potential  at low energies. At high energies this
model potential is insensitive to this parametrization  and leads to the
well-known Born-Oppenheimer form of exchange \cite{22}.

\section{Summary}

To summarize, we have used  a  time-reversal symmetric form of the
nonlocal model  
potential for exchange and applied it to the study  of Ps-H$_2$
scattering. 
We  have also 
presented  a simplified prescription for performing
coupled-channel dynamical scattering calculation 
with molecular target using a body-frame fixed-nuclei 
scheme.
With this prescription
 we have performed a three-Ps-state coupled-channel
 calculation of target-elastic Ps-H$_2$
scattering. Higher excitations and
ionization of the Ps atom are treated  using the first Born approximation  
with a regularized  exchange. We also calculated cross sections for 
two target-inelastic excitations
of H$_2$ (B ${}^1\Sigma_u^+$ and b ${}^3\Sigma_u^+$)
using first Born approximation with present exchange considering
Ps excitations 
($n=1,...,6$) and
ionization. 
Considering the fact that we have considered only two target-inelastic 
processes, 
the present total cross section is in encouraging agreement with
experiment.
The tractability of 
of the present dynamical calculational scheme for molecular targets  and 
the success of
the present time-reversal symmetric electron-exchange potential in
describing Ps-H$_2$ scattering should 
stimulate further investigation with both.

We thank  the Conselho Nacional de  Desenvolvimento Cient\'{\i}fico e
Tecnol\'ogico,  Funda\-\c c\~ao de Amparo \`a Pesquisa do Estado de S\~ao
Paulo,  and Financiadora de Estudos e Projetos of Brazil for partial
financial support.

%\newpage 

Figure Caption:
\vskip 1cm

1. Angle integrated target-elastic Ps-H$_2$ partial cross sections at
different positronium energies: elastic (solid line) and Ps(2s+2p) 
excitation (dashed-dotted line) from three-state coupled-channel model at
low energies interpolated to exchange Born at high energies, Ps($6 \ge
n\ge 3$) excitation (dashed-double-dotted line), Ps ionization (dashed
line) from exchange Born.

2. Angle integrated target-inelastic Ps-H$_2$ partial and total cross
sections at different positronium energies to H$_2$ (B $^1\Sigma_u^+$) 
state using the first Born approximation: Ps(1s)  (dashed-dotted line),
Ps(2s+2p) 
(dashed-double-dotted line), Ps($6>n>2$) excitation (dotted
line),
Ps ionization (dashed line), and total (full line).

3. Angle integrated target-inelastic Ps-H$_2$ partial and total cross
sections at different positronium energies to H$_2$ (b $^3\Sigma_u^+$) 
state using the first Born approximation: Ps(1s)  (dashed-dotted line),
Ps(2s+2p) 
(dashed-double-dotted line), Ps($6>n>2$) excitation (dotted
line),
Ps ionization (dashed  line), total (full line), total cross 
section with the nonsymmetric potential from ref. \cite{12}
(dashed-triple-dotted line).

4. Total Ps-H$_2$ cross section at different positronium energies:  total
target elastic cross section from figure 1 (dashed line), 
total target-elastic (from figure 1) plus target-inelastic (from figures
2 and 3)
cross section 
(solid line), 
experimental data
(solid circles from ref. \cite{2}, solid square from ref.  \cite{5}). 

\vskip 1cm

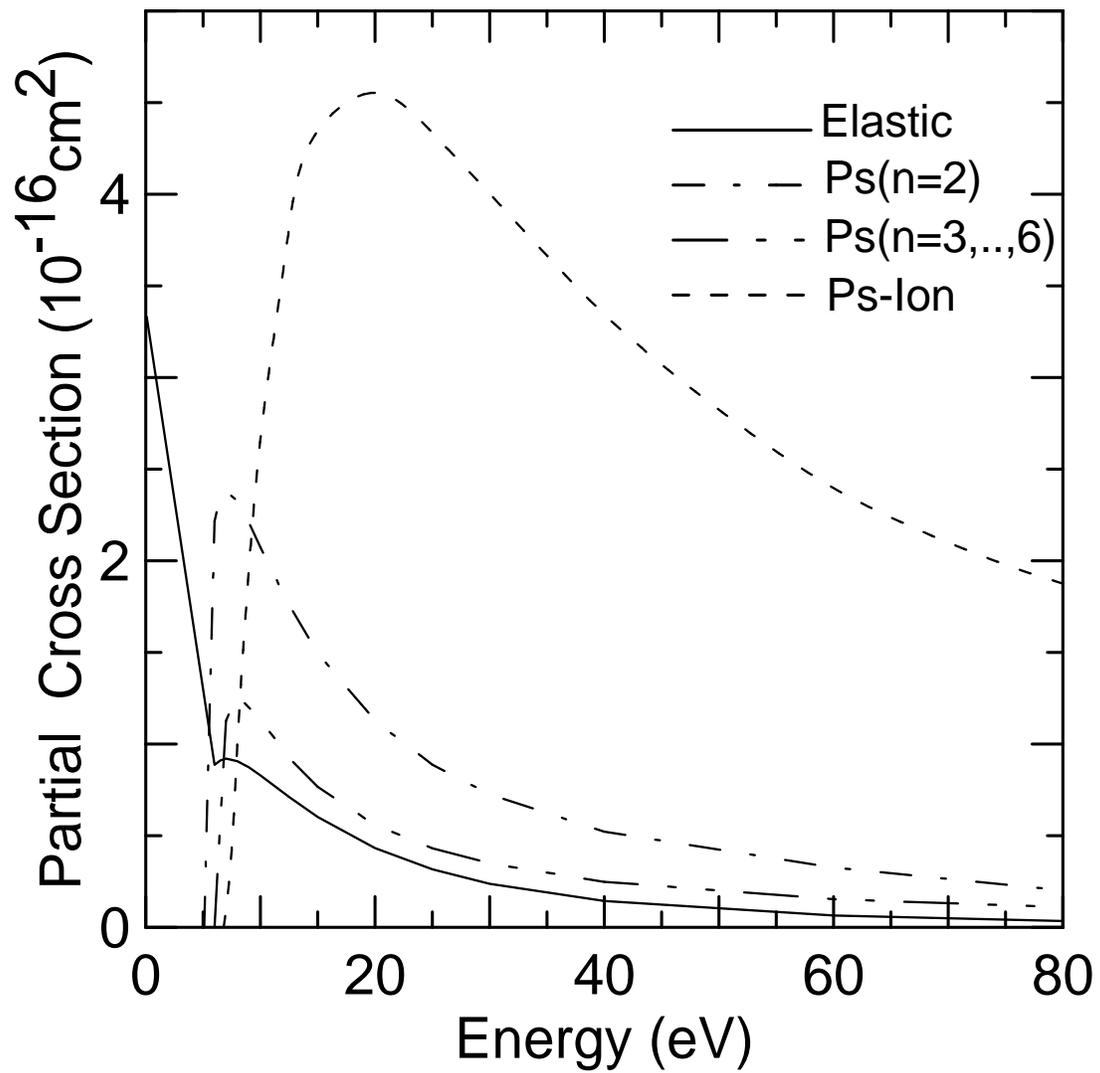
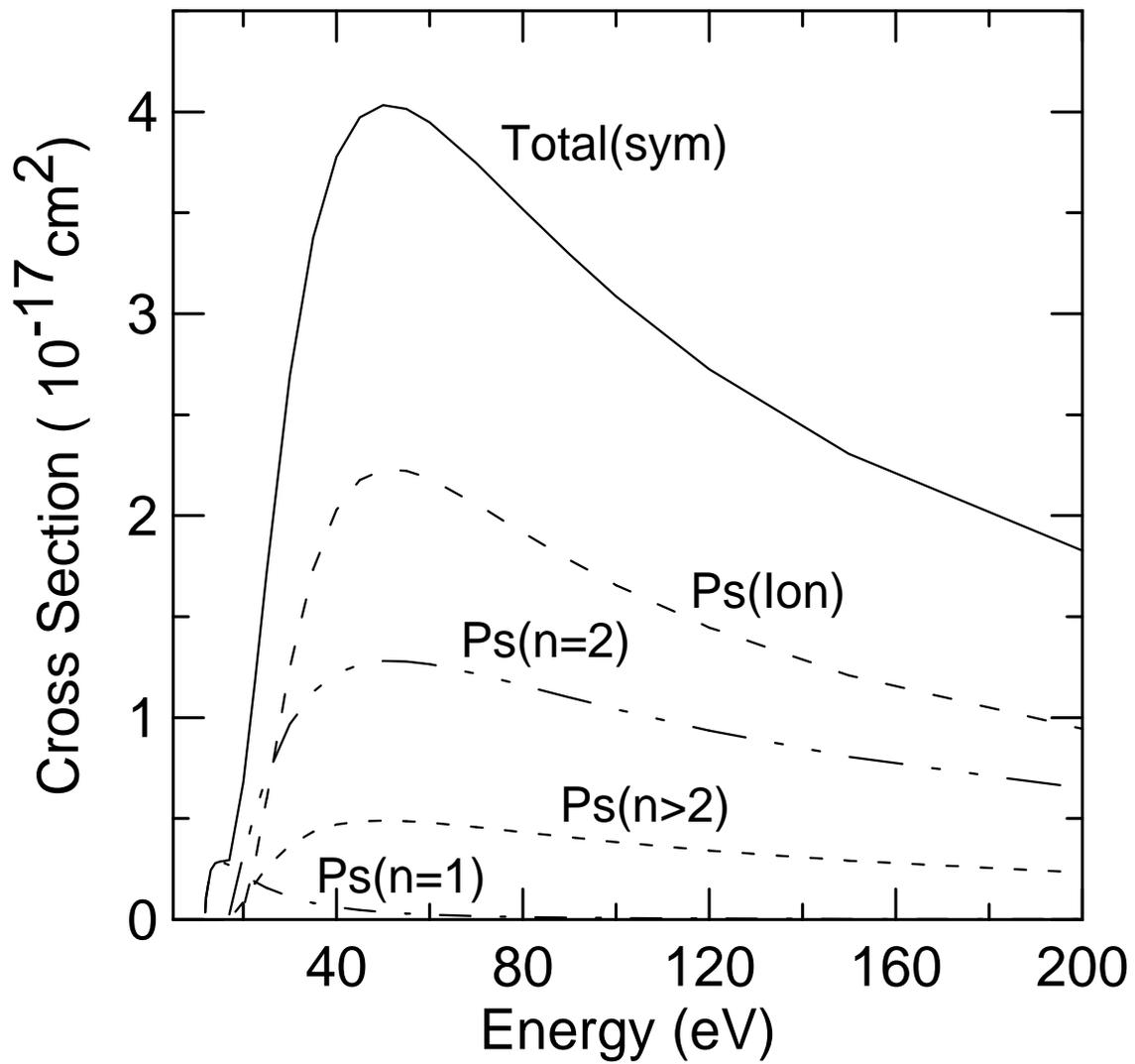
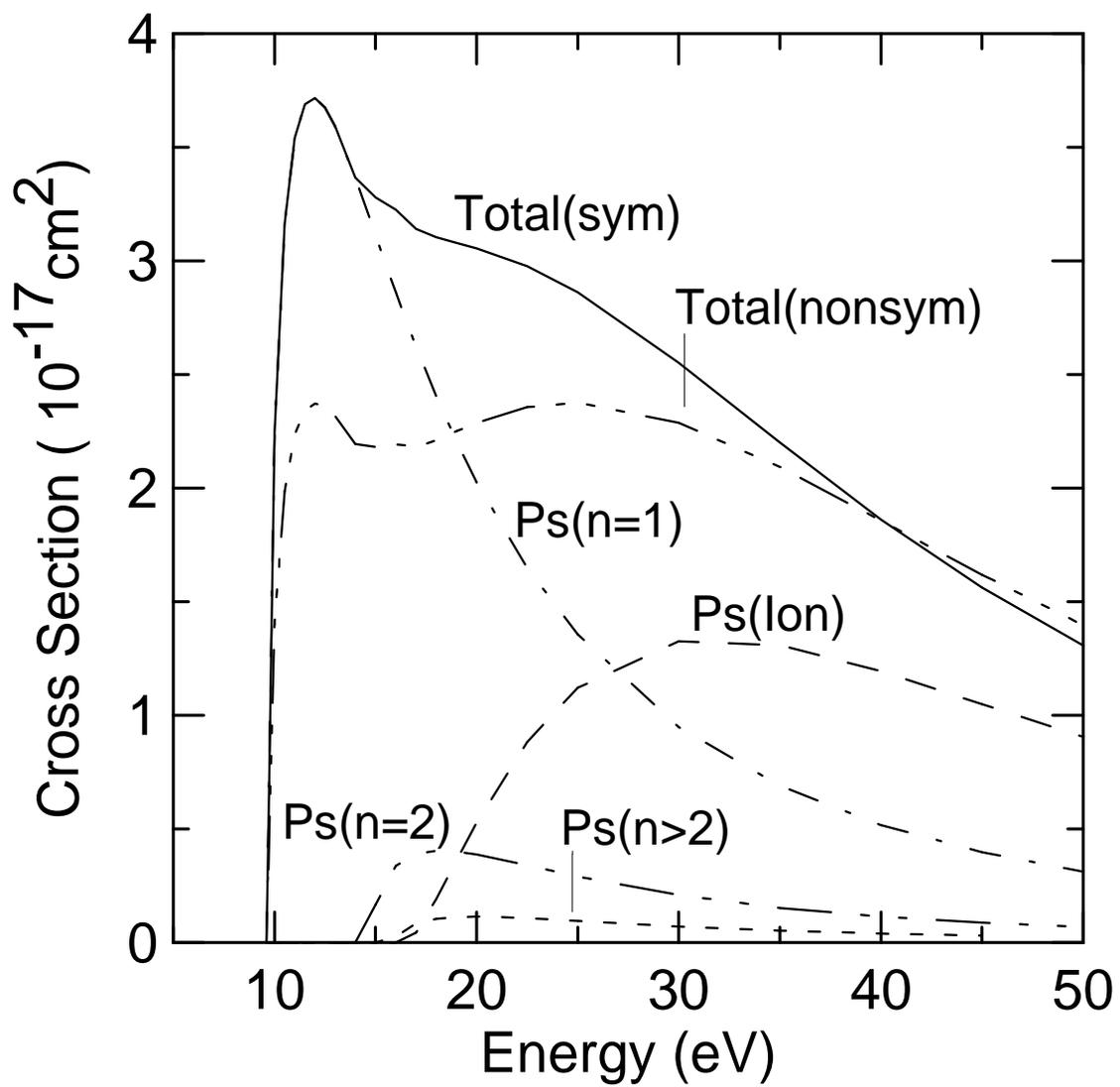
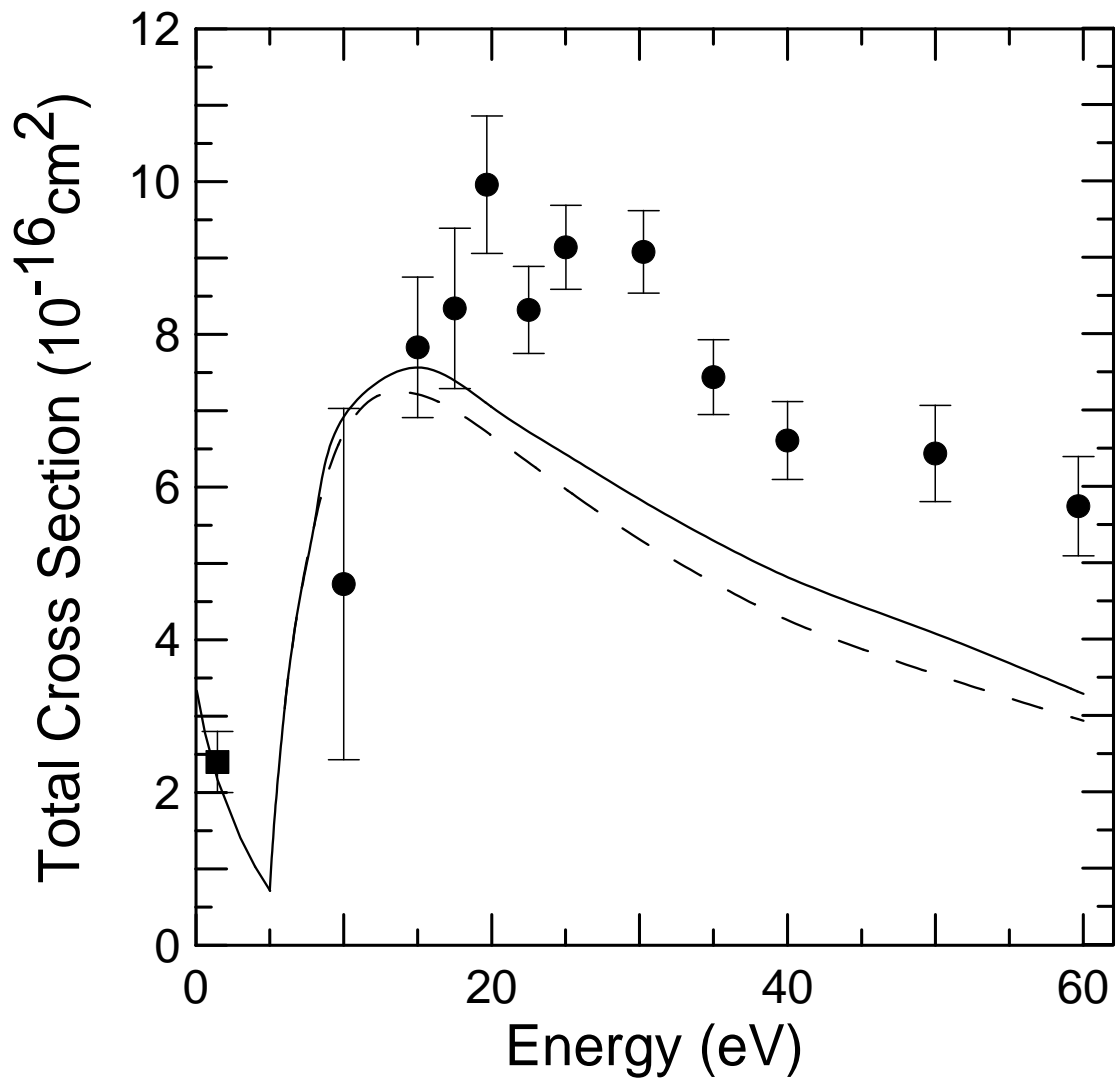
\end{document}